# Investigating the real-time dissolution of a compositionally complex alloy using inline ICP and correlation with XPS

Yao Qiu[1], Ruiliang Liu[1], Thomas Gengenbach[2], Oumaïma Gharbi[3], Sanjay Choudhary[1], Sebastian Thomas[1], Hamish L. Fraser[4] and Nick Birbilis[5]*

[1]Department of Materials Science and Engineering, Monash University, VIC 3800, Australia.
[2]CSIRO Manufacturing, Clayton, VIC, 3168, Australia.
[3]Laboratoire interfaces et systèmes électrochimiques, CNRS, Sorbonne Université, UMR8235, Paris, France.
[4]Department of Materials Science and Engineering, The Ohio State University, Columbus, OH. 43210, USA.
[5]College of Engineering and Computer Science, Australian National University, Acton, ACT 2601, Australia.

*nick.birbilis@anu.edu.au

**Abstract**

The real-time dissolution of the single-phase compositionally complex alloy (CCA), $Al_{1.5}TiVCr$, was studied using an inline inductively coupled plasma method. Compositionally complex alloys (CCAs), a term encompassing high entropy alloys (HEAs) or multi-principal element alloys (MPEAs), are - in general - noted for their inherently high corrosion resistance. In order to gain an insight into the dissolution of $Al_{1.5}TiVCr$ alloy, atomic emission spectroelectrochemistry was utilised in order to measure the ion dissolution of the alloy during anodic polarisation. It was revealed that incongruent dissolution occurred, with preferential dissolution of Al, and essentially no dissolution of Ti, until the point of alloy breakdown. Results were correlated with X-ray photoelectron spectroscopy, which revealed a complex surface oxide inclusive of unoxidised metal, and metal oxides in disproportion to the bulk alloying element ratio.

**Introduction**

The recent interest in high entropy alloys (HEAs)[1-3] has generated a large number of studies regarding alloys that are derived from the design concepts behind high entropy alloys. Notionally, HEAs are alloys with five or more alloy components in near equi-atomic proportions, and which present a high entropy of mixing, resulting in a single-phase structure. However, as is nowadays well documented, such compositionally complex alloys (CCAs) which are also termed multi-principal element alloys (MPEAs), are a much broader category of alloys that may or may not possess a high entropy of mixing, may not be single-phase, and may also have less than five principle alloying elements[4-7].

Of the works reported to date that have focused on the corrosion of HEAs, CCAs and MPEAs, there is, in the general case (notwithstanding exceptions) significant evidence that such alloys are highly corrosion resistant[3,8-11]. The high corrosion resistance of such alloys is an unintended, but welcome, consequence that is also coupled with alloy properties that include unprecedented hardness, and high strength[12,13]. In order to mechanistically study the corrosion of complex and corrosion resistant alloys, exposure testing and subsequent microscopy is not compatible with slow rates of alloy dissolution, therefore electrochemical testing is required in order to measure and quantify low levels of dissolution. Electrochemical testing, whilst extremely valuable (both critical and necessary) does not however offer any physical insight into the mechanism of corrosion or alloy dissolution, and requires supplementation with ancillary testing.

Of the most powerful analysis techniques now available to corrosion scientists and electrochemists, is the method developed by Ogle[14-16], termed atomic emission spectroelectrochemistry (AESEC). The AESEC method employs the use of an electrochemical flow cell, which permits standard electrochemical testing, however the flow cell is outlet into an inductively coupled plasma (ICP) instrument, which may either be ICP optical emission spectroscopy (ICP-OES)[14-18] or ICP mass spectroscopy (ICP-MS)[19,20]. This technique has been responsible for illuminating the dissolution of numerous alloy systems, by providing an element-selective insight whilst providing the precise rates of alloy dissolution – including for highly corrosion resistant alloys[20].

To the best of authors' knowledge, the AESEC method has not yet been applied to understanding the dissolution of CCAs. Of the analysis to date regarding corrosion of CCAs, including recent x-ray photoelectron spectroscopy analysis, it is posited that corrosion of CCAs involves incongruent dissolution – whereby there is selective oxidation of alloy species in proportions that are unique to the bulk alloy composition (and alloying element ratio)[11]. In fact, the complexity of the dissolution associated with CCAs was also recently posited as being so unique, that it challenges the long-held views of the thermodynamic nature of metal alloy passivity[21]. A first-order requirement of validating and quantifying incongruent dissolution associated with CCAs would further illuminate the process of alloy dissolution; but then also serve as a basis for illuminating second-order effects such as dissolution mechanisms and the nature of what passivity means in the cases of CCAs (or at least, for the alloy and its associated CCA family studied herein). The present study applies AESEC to the $Al_{1.5}TiVCr$ alloy, which is a close variant to the equi-atomic AlTiVCr alloy studied and reported in detail previously by Qiu and co-workers[4,11,21]. The alloy notation for $Al_{1.5}TiVCr$ indicates that the alloy is nominally comprised of equi-atomic proportions of Ti, V and Cr, but the atomic concentration of Al is 1.5 times that of Ti, V and Cr. This alloy is being studied because it represents an alloy that is single-phase and of appreciably low density (of 4.65 g/cm$^3$) in comparison to other CCAs reported to date; whilst also presenting high aqueous corrosion resistance relative to typical corrosion resistant alloys (CRAs) such as stainless steels (which nominally possess a density of > of 7.48 g/cm$^3$) or Ni alloys (which nominally possess a density of > of 8.09 g/cm$^3$).

**Results and Discussion**

Alloy microstructure

The $Al_{1.5}TiVCr$ alloy studied herein was investigated in the as-arc melted condition. In the as-arc melted condition, the $Al_{1.5}TiVCr$ alloy is single-phase, with the general microstructure typified by the electron backscatter diffraction inverse pole figure map presented in Figure 1a. Transmission electron microscopy (TEM) of $Al_{1.5}TiVCr$ in dark field mode, as shown in Figure 1b, reveals fine nanoscale domains (indicated by arrows) when imaged using the (100) superlattice reflection along the [001] zone axis. Accompanying selected area diffraction patterns collected from the [001] and [011] zone axis, reveal the characteristics of a uniform B2 structure (Figure 1c and 1d).

Atomic emission spectroelectrochemistry

The ion dissolution rate, represented in the form of an equivalent current density, for $Al_{1.5}TiVCr$ exposed to quiescent 0.1 M NaCl under open circuit conditions is presented in Figure 2a. It is observed that the method employed herein is not only capable of providing an equivalent current density, but the unique ion dissolution current density for individual species, namely, $i_{Al^{3+}}$, $i_{Ti^{4+}}$, $i_{V^{3+}}$, and $i_{Cr^{3+}}$ as determined in real time by online ICP-MS analysis. Therefore, the method employed is capable of providing ground truth dissolution rates on an elemental basis. The results in Figure 2a indicate that upon immersion, the dissolution of Al is greatest, particularly in the first ~250s of immersion, but then remaining as the principle dissolving element of the alloy. The dissolution of species at open circuit was observed to occur in the order of: $i_{Al^{3+}} \gg i_{V^{3+}} \gg i_{Cr^{3+}} > i_{Ti^{4+}}$, providing direct evidence for incongruent alloy dissolution in real time and in-situ.

The application of a polarising signal that is afforded by the atomic emission spectroelectrochemistry method, permits for the collection of a significant insight from the data presented in Figure 2b. The data shown in Figure 2b, includes the ion dissolution current for three distinct regimes within the test. The ion dissolution is shown for ~500s of open circuit conditioning, followed by the ion dissolution shown for the period of potentiodynamic polarisation (PDP), which also lasted ~500s, covering a potential range of ~ 1.5 V, at a potential scan rate 3 mV/s. It should be noted that the polarising signal commenced at a potential 100 mV below the open circuit potential and was scanned upwards through the corrosion potential and to increasing anodic potentials (and thus, the first ~100 mV of polarisation was cathodic). Finally, upon cessation of the applied polarisation signal, the ion dissolution rate is also shown for a period of open circuit (~200s).

From Figure 2b, the differentiation between the ion dissolution of alloying elements during open circuit is difficult to discern owing to the scale of the plotted data – however this is covered further below. The application of a polarisation is indicated by the signal denoted as $i_{pstat}$. It is noted that during the application of a polarising signal, the ion dissolution rate remains low, until about ~850s (corresponding to a potential of ~ +0.4 $V_{SCE}$). At this stage, a corresponding increase can also be seen in the signal denoted as $i_{tot}$. The $i_{tot}$ signal corresponds to the sum of all of $i_{Al^{3+}} + i_{Ti^{4+}} + i_{V^{3+}} + i_{Cr^{3+}}$. It is noted that at all times, the $i_{tot}$ signal remains lower than the $i_{pstat}$ signal, which is indicative that not all of the polarising signal is converted to ion dissolution; meaning that a surface film is developing during the anodic polarisation scan. The results in Figure 2b indicate that the principal source of the $i_{tot}$ signal is from $i_{Al^{3+}}$, and to a lesser extent, $i_{V^{3+}}$, $i_{Cr^{3+}}$ and $i_{Ti^{4+}}$. The relative proportions are however best discerned from a logarithmic plot, as discussed below. Finally, upon cessation of the polarising signal, it is noted that ion dissolution persists for some ~200s (which corresponds to the remaining period of assessment for the testing conducted herein), albeit with a diminishing ion dissolution signal. The diminishing of the ion dissolution signal indicates that 'repassivation' of the Al$_{1.5}$TiVCr occurred, following anodic polarisation. It is noteworthy that Ti$^{4+}$ is the fastest metal cation that reaches to a lower and stable ion dissolution current, suggesting the highest repassivation ability of Ti among those constitutional elements. It seems that the repassivation of Ti also contributes to lower ion dissolution current of other metal cations including V$^{3+}$, Cr$^{3+}$ and Al$^{3+}$ after 1100s.

Figure 2c reveals a portion of the AESEC data (as reported in Figure 2b corresponding to the period of potentiodynamic polarisation) presented in the familiar format of applied potential vs. current density, with the unique current density provided for $i_{Al^{3+}}$, $i_{Ti^{4+}}$, $i_{V^{3+}}$, $i_{Cr^{3+}}$, $i_{tot}$, and $i_{pstat}$. The representation in Figure 2c provides an additional insight into the dissolution of Al$_{1.5}$TiVCr. Whilst it was difficult from Figure 2b to discern the differences in ion dissolution between unique alloying elements, particularly at the early stages of polarisation, Figure 2c permits ready differentiation (owing to the logarithmic axis). The results in Figure 2c reveal that the dissolution is incongruent, not only with the principal ion dissolution associated with $i_{Al^{3+}}$, but there are definitive differences in the ion dissolution rates from $i_{V^{3+}}$, $i_{Cr^{3+}}$ and $i_{Ti^{4+}}$ (all of which are present in the alloy in equal proportions). The assessment of Figure 2 visually reveals the following: There are relatively uniform ion dissolution rates for all elements until ~ +0.3 V$_{SCE}$; The ion dissolution rates are unique depending on the element being dissolved, indicating incongruent dissolution and unique dissolution rates for all of the four elements in the alloy; At ~ +0.3 V$_{SCE}$, there is an increase in the $i_{Cr^{3+}}$ signal, and a corresponding decrease in the $i_{Al^{3+}}$ signal, At ~ +0.45 V$_{SCE}$, there is a sharp increase in the $i_{Al^{3+}}$ and the $i_{V^{3+}}$ signals, which when added to the increasing $i_{Cr^{3+}}$ signal, results in a marked increase in the $i_{tot}$ signal detected; Uniquely, the $i_{Ti^{4+}}$ signal begins to increase (by a marked 'breakdown') at ~ +0.7 V$_{SCE}$.

X-ray photoelectron spectroscopy (XPS) surface analysis of Al$_{1.5}$TiVCr

The abridged results from XPS characterisation of Al$_{1.5}$TiVCr are presented in Figure 3. The XPS results from native surface oxide upon Al$_{1.5}$TiVCr (naturally developed in laboratory air) are presented in Figures 3a and 3b.

The results in Figure 3a indicate the measured atomic concentrations as a function of surface depth, defined by etch time. As has been noted in prior works regarding XPS of a like alloy[21], the sputtering rates for CCAs are not (accurately) known and therefore sputtering time is not converted to depth. In the present study we do not report the detailed spectra, but the analysis of spectra based on the analysis method previously reported in Ref[21]. The results in Figure 3a indicate that for 0 min etching time (which corresponds to the outer surface of the oxide), there is the highest concentration of oxygen (O) and the principal alloying element detected is Al (followed by Cr, V and Ti). It is noted that during etching, the expectation of realising a metal-only surface was not observed, with the O concentration decreasing (to ~20 mins etching) and then remaining relatively consistent thereafter. This inability to etch to a wholly metal substrate, even under ultra-high vacuum, was previously discussed[21] and was attributed to the surface being highly reactive, immediately oxidising upon etching – even with the very minimal oxygen present in the XPS chamber (see Methods). Such a finding is characteristic of CCAs, and suggests that the origins of corrosion resistance of CCAs may not be in their low reactivity, but in high reactivity that leads to rapid oxidation and the development of protective surface films. It is noted, that XPS analysis was repeated numerous times for the samples discussed in this work, with highly consistent results (even when left in the XPS chamber for a conditioning for 3 days, in testing that is not reported herein). Based on the data format presented, it is notionally considered that the first few minutes of etching corresponds to the surface film propagation, with a transition through the film being completed within ~5 mins of etching, which is better evidenced by Figure 3b.

Figure 3b further analyses the data from the native surface film, differentiating the data in Figure 3a according to whether the species is an oxide ($M^{X+}$) or unoxidized metal ($M^0$). It is seen that there is an inhomogeneous surface oxide as a function of etching time. The XPS data from the outer surface of the alloy indicates that the highest proportion of metal species is present as an oxide (~60%). However, this proportion of oxide-to-unoxidised metal changes rapidly (within ~5mins of etching) with depth into the outer surface. In particular, for V and Cr, they are principally present as unoxidised metal (for the low proportion of their overall presence) in the surface film following ~5 mins of etching. At all times however, the proportion of Al oxide to Al metal remains > 1, such that (also when interpreted with Al having the highest surface concentration), the principle surface oxide is Al oxide with XPS confirms that the cation is predominantly $Al^{3+}$, doped with unoxidised metal from the alloying elements. Overall, all four metals exist at the surface (top few nm) in metallic (unoxidised metal, $M^0$) as well as oxidised form, and based on the experiments herein, the prevalent state is the most stable form ($Al^{3+}$, $Ti^{4+}$, $Cr^{3+}$ and $V^{3+}$). It is believed that an important feature of the data presented herein, and similarly in previous studies[21], is the nontrivial proportion of unoxidised metal ($M^0$) incorporated into the surface film – which is postulated as being a key contributor to the corrosion resistance of CCAs, although the mechanism is hitherto unexplored[11].

Figures 3c and 3d present the XPS results corresponding to the $Al_{1.5}TiVCr$ surface following anodic polarisation (as per Figure 2). Of note is that, following polarisation, the total atomic concentration of Al (Figure 3c) is lower at the outer surface of the alloy, and remains ~ 5-7 at. % lower through the surface. However, of principle difference between the XPS results for the native surface and the previously anodically polarised is the determination that outer surface film (Figure 3d) is principally comprised of oxides of the alloying elements (with negligible unoxidised metal) until ~5 mins of etching. Following ~5 mins of etching the proportion of oxide-to-unoxidised metal was similar to that of the native alloy surface film.

To investigate the re-oxidation behaviour of both the native surface and anodically polarised one, a "repeat" test was carried out after 60 mins of etching (see Figure 3). The native surface was briefly exposed to air, while the anodically polarised one was kept under vacuum for three days. It can be seen from both Figure 3b and 3d that there was a large increase in fraction of Ti oxide, V oxide and Cr oxide, however, a much lesser increase of Al oxide in both cases. It is interesting considering Al is the most reactive constitutional element to form oxides indicated by the most negative $\Delta G$ for metal oxidation ($-\Delta G^0 = -RTln\ pO_2$). It is likely due to the fact that Al was largely oxidised during sputtering (> 50 %). It is also possible related to the slow kinetics of Al oxide formation, such slow-growing of alumina ($Al_2O_3$) has been reported in Al-containing high temperature alloys and such slow-forming, dense oxide is beneficial for their oxidation property[22].

General discussion

The data in Figure 2, namely Figure 2c indicates that the dissolution of $Al_{1.5}TiVCr$ is incongruent. It is noted that each element has a unique dissolution rate, and a unique potential at which the unique ion dissolution rate increases with potential. It was also revealed that the $i_{tot} < i_{pstat}$ at all times during the AESEC testing, indicating the development of a surface film upon the alloy. The precise quantification of ion dissolution rates indicates the 'corrosion rate' of the alloy tested herein by a rapid in-situ methodology, also indicating that the alloy is capable of repassivation. However, it is important to bear in mind that one inherent limitation of this AESEC technique as the oxidation state of each element cannot be determined by ICP-MS. The ion dissolution rate of each element is calculated based on the most stable oxidation state of each element. In this study, CCA was polarised to a high potential (~1.0 $V_{SCE}$), there is a possibility of oxidation state change during polarisation, particularly for V and Cr. Despite the difficulty in empirical determination of actual oxide states, interpretation of the XPS analysis herein revealed that qualitatively, the fraction of $V^{4+}$ and $V^{5+}$ increased from around ~30% to ~60%, and the fraction of $Cr^{5+}$ was also increased and proportionally higher following polarisation. To even more accurately determine the ion dissolution rate, it might be useful to explore pure metals of numerous oxidation states using AESEC method, which will be important future work (and not carried out herein). The work herein is finite in its remit, but it has been the first demonstration of AESEC upon a CCA. The accompanying XPS results indicate that the CCA tested has a surface film that is principally comprised of the Al oxide, with the less reactive (or more noble) elements entrapped in the surface film, including in the unoxidised metallic state. As noted by a review of the corrosion of HEAs[3] and in works that have recently been studying the corrosion of HEAs (and CCAs)[21,23-29] there is important – and significant - future work required to rationalise the corrosion and surface films developed upon such complex alloys.

**Conclusions**

The present study has indicated that the dissolution of the alloy $Al_{1.5}TiVCr$ was in-congruent, both during open circuit exposure, and during potentiodynamic polarisation. The inline ICP-MS method employed, termed atomic emission spectroelectrochemistry, was capable of providing significant insights into the alloy dissolution mechanism. It was revealed that $Al^{3+}$ dissolution was 1 - 2 orders of magnitude greater than V, Cr and Ti, respectively. In that context, the significant majority of the total dissolution current ($i_{tot}$) measured was from Al dissolution. The measured $i_{tot}$ was lower than the measured potentiostat current, $i_{pstat}$, indicative of the development of a surface film upon the alloy during dissolution. Upon cessation of the applied anodic potential, the real-time repassivation of $Al_{1.5}TiVCr$ was readily observed. The aforementioned surface film developed on $Al_{1.5}TiVCr$ was studied using XPS. The XPS analysis of the native surface oxide, and that following polarisation were notionally consistent, revealing a low proportion of Ti and the dominance of Al in the film. The predominant oxide in the surface film was that of oxidised Al, and to a significantly lesser extent, Cr, V and then Ti. There was evidence that there was unoxidised metal in the surface film, previously posited in prior work, with evidence of unoxidised metal for all of Al, Ti, Cr and V.

**Data availability**

All relevant data are available from authors upon reasonable request.

**Acknowledgements**

We thank the Monash Centre for Electron Microscopy (MCEM). Dr. Mark Gibson is also gratefully acknowledged for technical assistance with alloy production.

**Author contribution**

Yao Qiu: Designed the alloy, did all the electron microscopy, coordinated this study. Ruiliang Liu: Acquisition and processing of ICP-MS data. Thomas Gengenbach: Acquisition, processing and evaluation of XPS data. Oumaïma Gharbi: Technical support with ICP-MS. Oumaïma Gharbi, Sanjay Choudhary, Sebastian Thomas, Hamish L. Fraser: Commented on the paper. Nick Birbilis: Supervised this study. Nick Birbilis and Yao Qiu: Wrote the paper.

**Additional information**

Competing interests: The authors declare no Competing Financial or non-Financial Interests.

**Methods**

Alloy production

The $Al_{1.5}TiVCr$ alloy was produced by arc melting in an argon atmosphere, with pure (99.99 wt. %) metal pieces (chunks) as the starting materials.

Microstructural characterisation

Electron backscatter diffraction (EBSD) of the metallographically prepared $Al_{1.5}TiVCr$ surface was carried out using an FEI Quanta 3D FEG, equipped with a Pegasus Hikari EBSD System and using TSL software. Specimen preparation involved gradually polishing to 1 μm finish using diamond paste following by final polishing with 0.05 μm oxide polishing suspension (OPS).

Transmission electron microscopy (TEM) was carried out using an FEI Tecnai T20 instrument. Specimen preparation for TEM was carried out using the focused ion beam (FIB) lift-out technique, in order to create an electron transparent lamella.

AESEC

Atomic emission spectroelectrochemical (AESEC) measurements involved the use of a scanning flow cell coupled with an inductively coupled plasma mass spectrometer (ICP-MS). A conventional three-electrode configuration was employed for electrochemical testing, with such a configuration realised using a commercially available electrochemical flow cell (012799, ALS Co., Ltd, Tokyo, Japan). Unbuffered quiescent 0.1 M NaCl was used as the electrolyte, which was continuously pumped past the working electrode surface and injected onto the ICP-MS instrument using a flow rate of ~0.13 mL/min, allowing for real-time in line elemental analysis of the following elements $^{27}$Al, $^{52}$Cr, $^{48}$Ti, $^{51}$V. Electrochemical testing was carried out using a potential scan rate of 3 mV/s, and potentiodynamic polarisation was carried out in an upward scan direction from the open circuit potential (OCP) −100 mV$_{SCE}$. A Bio-logic SP-50 potentiostat was used for the electrochemical control during AESEC. The ICP-MS instrument used in this study was a Perkin Elmer Optima 8000 ICP-MS Spectrometer using Syngistix software. The ICP-MS instrument detection limit (C$_{2\sigma}$) for the following element Al, Cr, Ti, V was 3, 2, 0.5, and 0.5 µg/L.

The AESEC method has a high sensitivity to the detection of dissolved ion concentration, as well as micrometer-sized solids (precipitates or particles) during the process of open circuit exposure or polarisation of metallic alloys. The calculation of the concentration of released metal ions and particles as a function of time has been presented in numerous studies[14,30,31] and as per prior works, the instantaneous metal dissolution rate was calculated (expressed in µg/s cm$^2$) using the following equation:

$$v_M = C_M f / A$$

With C$_M$ the volume adjusted ion concentration (µg/L), $f$ the electrolyte flow rate (mL/s) and A the exposed working electrode surface area (0.6 cm$^2$)[31]. The conversion of the instantaneous dissolution rate to the partial elemental current density (i$_M^{X+}$) was determined *via* Faraday's law.

XPS analysis

X-ray photoelectron spectroscopy (XPS) was carried out using an AXIS Nova spectrometer (Kratos Analytical, UK) with a monochromated Al K$\alpha$ source, and utilising the standard aperture (0.7 mm × 0.3 mm). Survey spectra were acquired at a pass energy of 160 eV. Higher resolution spectra were recorded from peaks at 40 eV pass energy, and binding energies were referenced to the Ti2p$_{3/2}$ metal peak at 453.9 eV analysed at an emission angle of 0 degrees. The reported depth profiling experiments were conducted using an Ar Gas Cluster Ion Source with a cluster size of Ar1000$^+$ and impact energy of 10 keV. The total vacuum chamber pressure was ~10$^{-9}$ mbar, and the residual gas was analysed using an MKS Instruments e-Vision 2 quadrupole mass spectrometer.

**Figures**

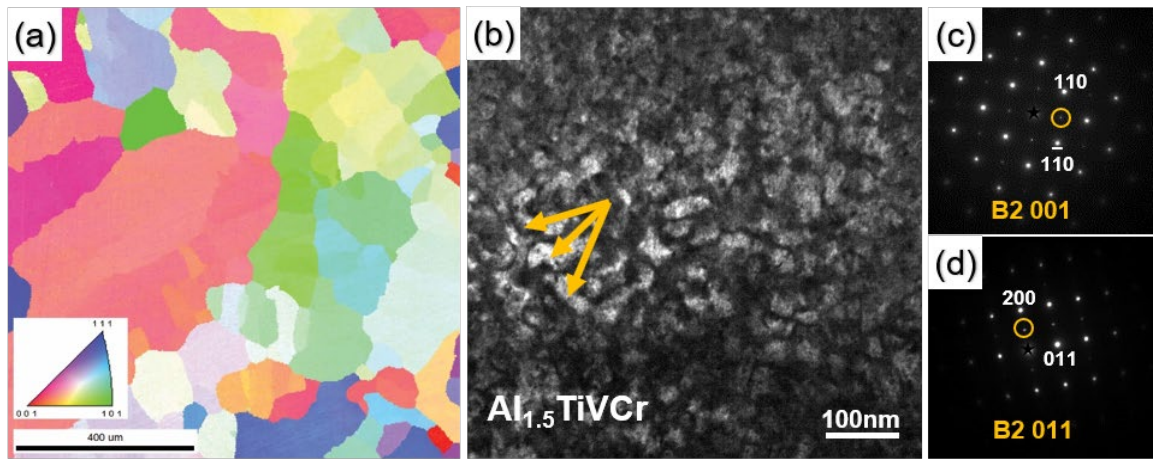

**Figure 1**: (a) Microstructure of as-cast single-phase Al$_{1.5}$TiVCr as represented from an EBSD inverse pole figure map, (b) Dark field TEM image obtained using the (100) superlattice reflection as determined from the selected area electron diffraction (SAED) pattern along the [001] zone axis (revealing the extremely fine domains (indicated by arrows)). (c-d) SAED patterns of the alloy along [001] and [011] zone axis, revealing characteristics of a uniform B2 structure.

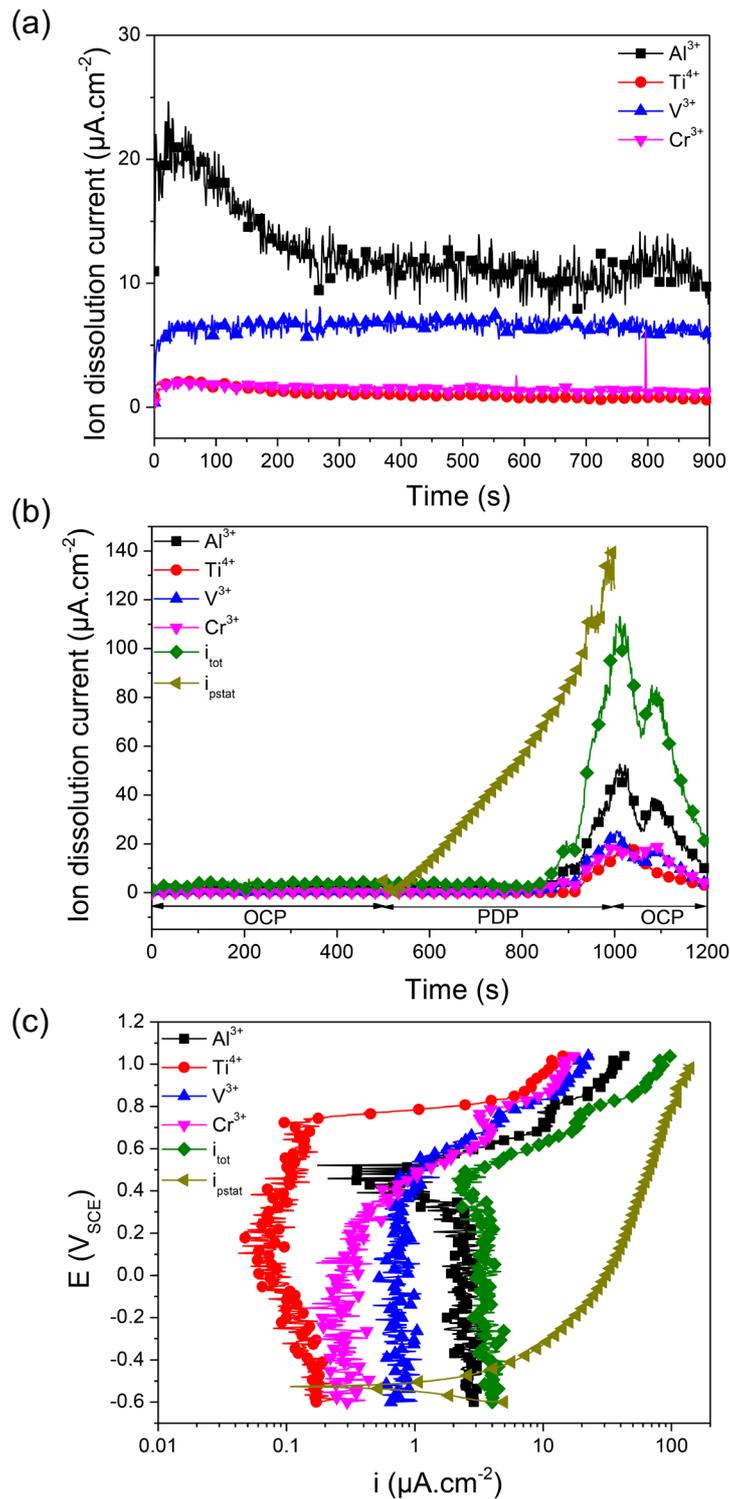

**Figure 2**: Current density $i_{Al^{3+}}$, $i_{Ti^{4+}}$, $i_{V^{3+}}$, $i_{Cr^{3+}}$ determined by online ICP-MS analysis of $Al_{1.5}TiVCr$ in quiescent 0.1 M NaCl, during: (a) Open circuit potential (OCP) exposure for 15 mins; (b) open circuit condition and subsequent potentiodynamic polarisation exposure for the alloy produced herein in 0.1 M NaCl. The sum of ICP determined dissolution current $i_{tot}$ and the potentiostat applied current density $i_{pstat}$ during PDP exposure are also included for comparison purpose. (c) Potentiodynamic polarisation data presented in the format of applied potential vs. Current density $i_{Al^{3+}}$, $i_{Ti^{4+}}$, $i_{V^{3+}}$, $i_{Cr^{3+}}$, $i_{tot}$, and $i_{pstat}$.

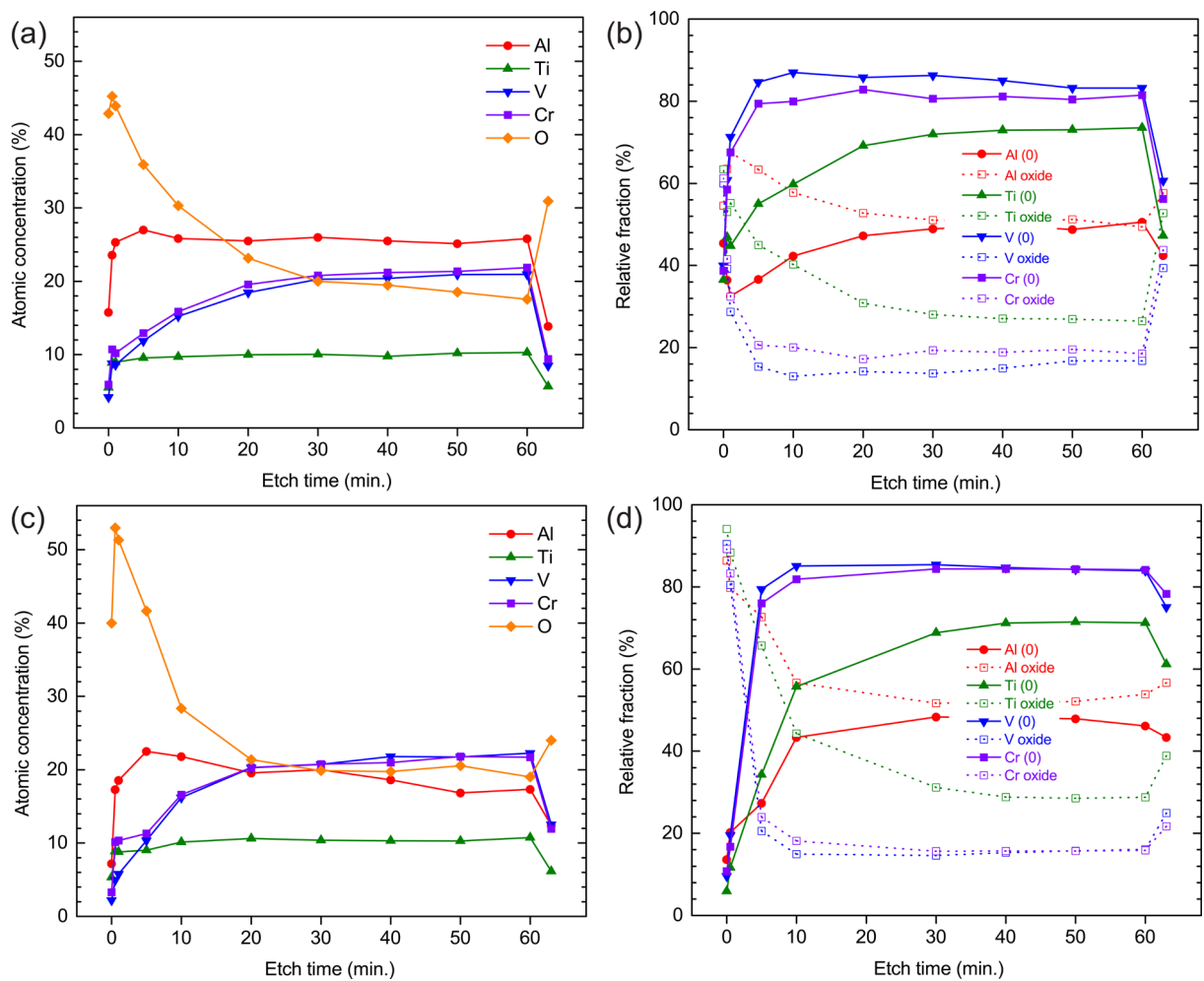

**Figure 3**: Data from XPS depth profiles collected for: (a) native surface oxide upon Al$_{1.5}$TiCrV; (b) data accompanying the results in (a) for the native surface oxide upon Al$_{1.5}$TiCrV, however processed to provide the relative fraction of oxide to unoxidised metal (M$^0$); (c) the surface film upon Al$_{1.5}$TiCrV following an anodic potentiodynamic polarisation scan (as per Fig. 2c); (d) data accompanying the results in (c) for the surface oxide following anodic polarisation upon Al$_{1.5}$TiCrV, however processed to provide the relative fraction of oxide to unoxidised metal (M$^0$). Note that etching stopped at 60 min. The final data point (plotted at just above 60 min. for comparison) was measured after the samples had either been exposed very briefly to air (3(a) and (b)) or left under vacuum for three days (3(c) and (d)). See text for details.